\documentclass[
 twocolumn,
 superscriptaddress,
 amsmath,
 amssymb,
 floatfix,
 longbibliography,
 balancelastpage]{revtex4-2}

\usepackage[resetlabels,labeled]{multibib}
\usepackage[utf8]{inputenc}

\usepackage{amsthm,graphics,bbm}
\usepackage{epsfig}
\usepackage{booktabs}

\usepackage{graphicx}
\usepackage{dcolumn}
\usepackage{bm}
\usepackage{xfrac}



\def\be{\begin{equation}}
\def\ee{\end{equation}}
\def\bea{\begin{eqnarray}}
\def\eea{\end{eqnarray}}

\usepackage{siunitx}
\usepackage{color}
\usepackage{comment}

\newcommand{\av}[1]{\left\langle#1\right\rangle}

\usepackage[dvipsnames]{xcolor}
\usepackage[hyperindex,breaklinks,colorlinks,citecolor=blue,urlcolor=black,linkcolor=black]{hyperref}
\usepackage{ulem}
\usepackage{soul}

\begin{document}

\preprint{APS/123-QED}

\title{
Measuring Hall voltage and Hall resistance in an atom-based quantum simulator
}

\affiliation{Department of Physics and Astronomy, University of Florence, 50019 Sesto Fiorentino, Italy}
\affiliation{European Laboratory for Non-Linear Spectroscopy (LENS), 50019 Sesto Fiorentino, Italy}
\affiliation{Istituto Nazionale di Ottica del Consiglio Nazionale delle Ricerche (CNR-INO), Sezione di Sesto Fiorentino, 50019 Sesto Fiorentino, Italy}
\affiliation{Université Grenoble Alpes, CNRS, LPMMC, 38000 Grenoble, France}
\affiliation{Department of Quantum Matter Physics, University of Geneva, 1211 Geneva, Switzerland}
\affiliation{Université Grenoble Alpes, CEA, IRIG-MEM-L\_SIM, 38000 Grenoble, France}

\author{T.-W. Zhou}
\affiliation{Department of Physics and Astronomy, University of Florence, 50019 Sesto Fiorentino, Italy}

\author{T. Beller}
\affiliation{Department of Physics and Astronomy, University of Florence, 50019 Sesto Fiorentino, Italy}

\author{G. Masini}
\affiliation{European Laboratory for Non-Linear Spectroscopy (LENS), 50019 Sesto Fiorentino, Italy}

\author{J. Parravicini}
\affiliation{Department of Physics and Astronomy, University of Florence, 50019 Sesto Fiorentino, Italy}
\affiliation{European Laboratory for Non-Linear Spectroscopy (LENS), 50019 Sesto Fiorentino, Italy}
\affiliation{Istituto Nazionale di Ottica del Consiglio Nazionale delle Ricerche (CNR-INO), Sezione di Sesto Fiorentino, 50019 Sesto Fiorentino, Italy}

\author{G. Cappellini}
\affiliation{Istituto Nazionale di Ottica del Consiglio Nazionale delle Ricerche (CNR-INO), Sezione di Sesto Fiorentino, 50019 Sesto Fiorentino, Italy}
\affiliation{European Laboratory for Non-Linear Spectroscopy (LENS), 50019 Sesto Fiorentino, Italy}

\author{\\C. Repellin}
\affiliation{Université Grenoble Alpes, CNRS, LPMMC, 38000 Grenoble, France}

\author{T. Giamarchi}
\affiliation{Department of Quantum Matter Physics, University of Geneva, 1211 Geneva, Switzerland}

\author{J. Catani}
\affiliation{Istituto Nazionale di Ottica del Consiglio Nazionale delle Ricerche (CNR-INO), Sezione di Sesto Fiorentino, 50019 Sesto Fiorentino, Italy}
\affiliation{European Laboratory for Non-Linear Spectroscopy (LENS), 50019 Sesto Fiorentino, Italy}

\author{M. Filippone}
\affiliation{Université Grenoble Alpes, CEA, IRIG-MEM-L\_SIM, 38000 Grenoble, France}

\author{L. Fallani}
\email{fallani@lens.unifi.it}
\affiliation{Department of Physics and Astronomy, University of Florence, 50019 Sesto Fiorentino, Italy}
\affiliation{European Laboratory for Non-Linear Spectroscopy (LENS), 50019 Sesto Fiorentino, Italy}
\affiliation{Istituto Nazionale di Ottica del Consiglio Nazionale delle Ricerche (CNR-INO), Sezione di Sesto Fiorentino, 50019 Sesto Fiorentino, Italy}


\begin{abstract}
In the Hall effect, a voltage drop develops perpendicularly to the current flow in the presence of a magnetic field, leading to a transverse Hall resistance. Recent developments with quantum simulators have unveiled strongly correlated and universal manifestations of the Hall effect. However, a direct measurement of the Hall voltage and of the Hall resistance in a non-electronic system of strongly interacting fermions was not achieved to date. Here, we demonstrate a technique for measuring the Hall voltage in a neutral-atom-based quantum simulator. From that we provide the first direct measurement of the Hall resistance in a cold-atom analogue of a solid-state Hall bar and study its dependence on the carrier density, along with theoretical analyses. Our work closes a major gap between analogue quantum simulations and measurements performed in solid-state systems, providing a key tool for the exploration of the Hall effect in highly tunable and strongly correlated systems.  
\end{abstract}


\maketitle

\noindent The Hall effect~\cite{ashcroft1976solid,RevModPhys.58.519,prange1989quantum,Yoshioka} is a macroscopic manifestation of the Lorentz force exerted by a magnetic field on the charge carriers of a conductor. In normal conductors threaded by a magnetic field $B$, the Hall effect manifests itself as a voltage drop corresponding to the emergence of an electric field $E_y$, perpendicular to the current density flow $j_x$. The ratio between these two quantities defines the Hall resistivity
\begin{equation}\label{eq:RH}
\rho_{\rm H}=\frac{E_y}{j_x}\,.
\end{equation}
Despite its simple origin, the study of this quantity has played a central role for the exploration and understanding of solid-state systems~\cite{popovic_hall_2003,PhysRevLett.122.200001}. For small magnetic fields, within semiclassical approximations, the Hall resistivity $\rho_{\rm H}$ is proportional to $- B/nq$ and measures the inverse of the carrier density $n$ and their charge $q$. The observation of a negative Hall resistivity has been an early signature of electrical conduction by holes before the advent of band theory~\cite{ashcroft1976solid}. The study of the Hall effect under the action of strong magnetic fields has led to the discovery of the integer~\cite{klitzing1980new} and fractional~\cite{tsui_FQHE} quantum Hall effects, where $1/\rho_{\rm H}$ is exactly quantized in units and fractions of the universal constant $e^2/h$. The study of such phenomena has subsequently fostered the entire field of topological quantum matter~\cite{hasan2010colloquium,xiao2010_berry_phase_review,RevModPhys.88.035005,haldane2017nobel,wen2019choreographed}. 

\begin{figure}[t!]
    \centering
    \includegraphics[width=1\columnwidth]{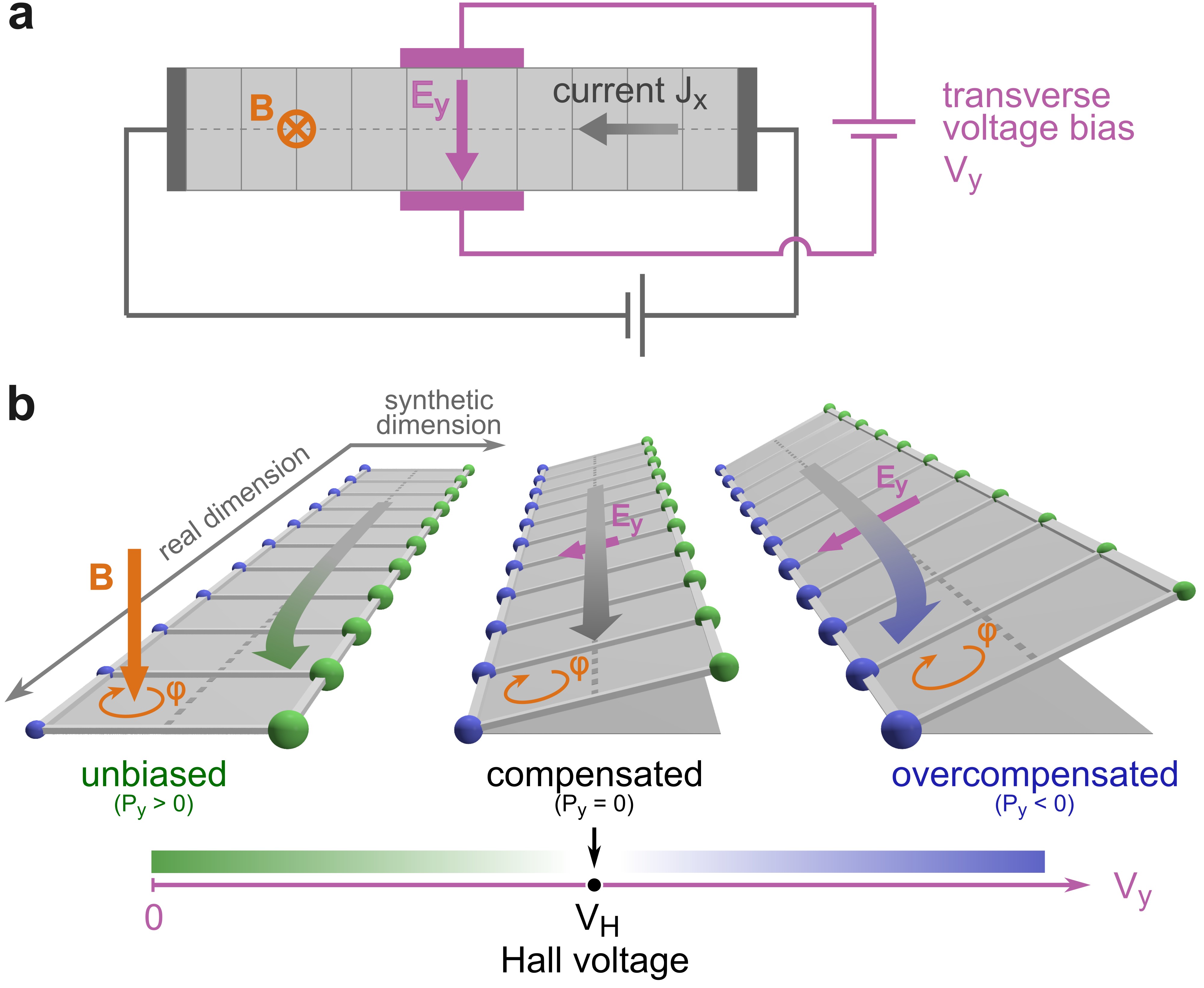}
    \caption{\textbf{Measuring electrical quantities in a synthetic Hall bar.} {\textbf{a}}, An analogue quantum simulation of a Hall bar is realized by trapping ultracold atoms in ladder geometries. An atomic current $J_x$ is subjected to a perpendicular effective magnetic field $B$ and to synthetic electric fields realized by energy gradients along both the longitudinal direction ($\hat{x}$) and the transverse direction ($\hat{y}$, encoded in the internal spin state according to the concept of synthetic dimension). {\textbf{b}}, The transverse voltage bias that compensates the bending of the carrier trajectories induced by the Hall effect (measured by the spin polarization $P_y$) provides a direct measurement of the Hall voltage $V_{\rm H}$, from which the Hall resistance is extracted.
    }
    \label{fig1}
\end{figure}

However, understanding the Hall effect has remained challenging whenever interactions are present among the carriers. The quest for further insight~\cite{goldman2016topological,Ozawa2019,xiang2023simulating} has motivated an intense experimental work in the last decade, with the realization of analogue quantum simulators exploiting ultracold atoms subjected to artificial magnetic fields~\cite{PhysRevLett.111.185301,Aidelsburger2015,PhysRevLett.111.185302,science.aaa8736,science.aaa8515,Greiner2017,genkina2019imaging,Mukherjee2022,viebahn2023,Greiner2023,zhou2023observation,Lunt2024,impertro2024realization} and circuit quantum electrodynamics~\cite{xiang2023simulating,wang2024realization}.


Recent experiments with cold atoms~\cite{chalopin_probing_2020,zhou2023observation} have measured the transverse polarization in synthetic flux ladders after inducing a transient longitudinal current, revealing a universal behaviour above an interaction threshold~\cite{zhou2023observation,greschner2019universal}. A precursor signature of the topological robustness of $\rho_{\rm H}$ for insulating phases has also been probed through St\v reda's formula~\cite{Greiner2023}. In normal conductors, the robustness of the connection between the Hall resistance and the inverse carrier density is deeply rooted in Galilean invariance~\cite{girvin2002quantum,lopatin2001hall}. However, despite being a natural quantity of easy access in the solid state, the transport measurement of the Hall resistance in a neutral system such as an atomic quantum simulator has remained out of reach so far. Such a measurement would permit the unprecedented investigations of the Hall response of strongly correlated conductors, bringing new input to the theoretical challenge~\cite{RevModPhys.91.011002,PhysRevB.4.1566,PhysRevB.75.195123,PhysRevLett.121.066601,zotos2000reactive,PhysRevB.55.3907,PhysRevLett.70.2004,PhysRevB.91.024507,citro2024hall} of understanding anomalous temperature dependence and sign changes of $\rho_{\rm H}$ in correlated solid-state systems~\cite{hagen_anomalous_1990,badoux_change_2016,smith_sign_1994,PhysRevLett.84.2670,PhysRevLett.84.2674}.

In this work, we demonstrate a quench protocol to measure the Hall resistance in a Hall bar of strongly interacting ultracold fermions, which generalizes the theoretical protocol of Ref.~\cite{PhysRevLett.126.030501} to finite magnetic fields. Our method is based on the tracking of the transverse polarization along the bar, and on the application of a controlled effective voltage bias in the transverse direction to compensate it. By controlling the number of particles in the system and the transverse size of the system, we provide direct experimental evidence of the predicted $1/n$ dependence of the Hall resistance~\eqref{eq:RH} on the carrier density $n$, irrespective of the microscopic details of the system. 
We also provide the theoretical demonstration that such behaviour is stabilized by strong repulsive interactions in ladder systems and that it is a manifestation of the universal Hall response of single-band metals~\cite{zhou2023observation,greschner2019universal}. Our experiment reproduces the electrical measurements performed on solid-state Hall devices and lays the foundations for the investigation of the quantum Hall effect and, more generally, of topologically protected transport in strongly correlated systems.

\vspace{1em}
\noindent\textbf{\large Results}

\begin{figure}[t!]
    \centering
    \includegraphics[width=\columnwidth]{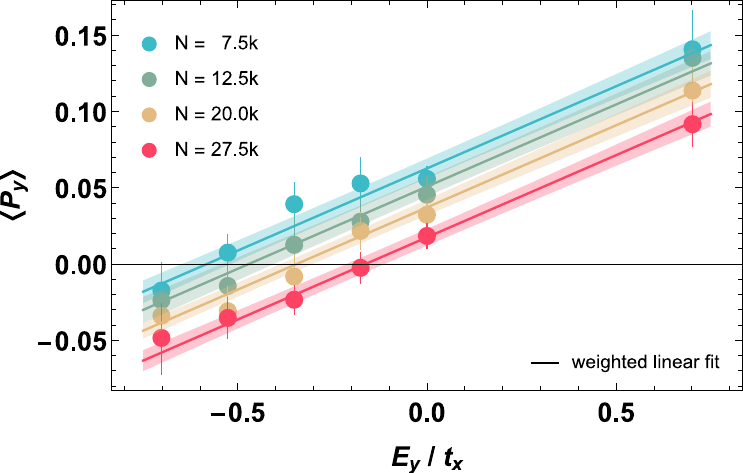}
    \caption{\textbf{Measurement of the Hall voltage.} The time-averaged Hall polarization $\langle P_y\rangle$ for two-leg ladders, measured at $t_y=3.30t_x$ and $U=6.56t_x$, is shown as a function of the transverse field $E_y$ with different total atom number $N$; the error bars depict the standard error of the mean and are obtained with a statistical Bootstrap method. The solid lines in corresponding colours are obtained from a weighted global linear fit based on the error bars of the experimental data, while colour shades represent the 95$\%$ confidence bands of the fit.}
    \label{fig2}
\end{figure}

\noindent{\bf Quantum simulation of a Hall bar -- }
Our experiment operates with ultracold Fermi gases of $^{173}$Yb atoms trapped in a deep 2D optical lattice, resulting in an array of fermionic 1D quantum wires (in the following called ``tubes'') along the $\hat{x}$-direction. We engineer Hall bars in the form of synthetic ladders, where the dynamics along the longitudinal direction is controlled by an additional optical lattice along $\hat{x}$.
The nuclear-spin states of the atoms act as different sites along a synthetic dimension $\hat{y}$, which provides the transverse direction of the ladder~\cite{science.aaa8736,zhou2023observation}, see Fig.~\ref{fig1}. The plaquettes of the ladder are threaded by a synthetic magnetic flux $\varphi$, generated by phase imprinting by Raman lasers coupling the different spin states, which simulates the action of a magnetic field $B$ perpendicular to the Hall bar (see Methods for details about the experiment).
We realize two- and three-leg ladder systems where a current $J_x$ along the longitudinal $\hat x$-direction is induced by an energy gradient along the same direction, which is equivalent to applying a longitudinal electric field $E_x$ (see Fig.~\ref{fig1}). The Hall effect is signalled by the emergence of a transverse density polarization of the system $P_y$~\cite{greschner2019universal,filippone2019vanishing}. Here we directly measure the transverse Hall voltage $V_{\rm H}$ caused by this polarization, by applying an energy offset between the legs of the ladder, so as to mimic the effect of a transverse electric field $E_y$ (see following section), and adjusting it such that $P_y=0$, see Figs.~\ref{fig1} and~\ref{fig2}. Such a measurement of $E_y$ allows us to derive the reactive Hall resistance~\cite{prelovsek1999hall,zotos2000reactive,PhysRevLett.126.030501}.

The system, prior to the quench protocol of activating the longitudinal current and the transverse field, can be described by the interacting Harper-Hofstadter Hamiltonian 
\begin{align}\label{eq1}
H=&-t_x\sum_{j,m}\left[a^{\dagger}_{j,m}a^{\phantom \dagger}_{j+1,m}+\textrm{h.c.}\right]+\frac{U}{2}\sum_{j,m,m'\neq m}n_{j,m}n_{j,m'}\nonumber\\
&-t_y\sum_{j,m}\left[e^{i\varphi j}a^\dagger_{j,m}a^{\phantom \dagger}_{j,m+1}+\textrm{h.c.}\right]\,,
\end{align}
where $a^{\phantom \dagger}_{j,m}$ and $a^{\dagger}_{j,m}$ are fermionic annihilation and creation operators acting on site $(j,m)$, $n_{j,m}=a^\dagger_{j,m}a^{\phantom\dagger}_{j,m}$ and h.c. is the Hermitian conjugate operator. The lattice label $j$ corresponds to the real $\hat x$-dimension, while the labels $m,\,m'\in\lbrack1,M\rbrack$ indicate the synthetic $\hat y$-dimension with $M=2$ or $3$, corresponding to the nuclear spin states $m_{\rm F}=-\sfrac52,\,-\sfrac12$ and $+\sfrac32$. 
Here, $t_x$ is the nearest-neighbour tunneling amplitude and $U$ 
is the ``on-rung'' repulsive interaction energy between two atoms with different nuclear spin in the same real-lattice site and with global SU($M$) interaction symmetry~\cite{RN2022}. The dynamics along the synthetic dimension is encoded in the complex hopping amplitude $t_y e^{i\varphi j}$, whereby the position-dependent phase simulates the effect of a static magnetic flux $\varphi=0.32\pi$ threading a plaquette of the ladder (see Methods).

In the non-interacting limit ($U=0$), the single particle spectrum of Eq.~\eqref{eq1} is composed of $M$ independent conduction bands. However, the strong repulsion among the nuclear spin states stabilizes a single-band correlated metal, whose hallmark in the two-leg case was shown to be a universal Hall imbalance $\Delta_{\rm H}={P_y}/{J_x}=2t_x\tan(\varphi/2)/t_y$, robust to variations of particle densities, confinement, finite temperatures and strong drives~\cite{zhou2023observation}. We will show that such behaviour in the universal single-band regime underpins the observation of the universal scaling with $1/n$ of the Hall resistance~\cite{greschner2019universal,PhysRevLett.126.030501}, and also extends to a three-leg ladder.

\vspace{0.5em}
\noindent{\bf Experimental protocol -- }
The experiment is performed in the universal single-band regime at $t_y=3.30t_x$ and $U=6.56t_x$~\cite{zhou2023observation}. 
A light-shift gradient, which results in the additional Hamiltonian term $H_x=E_x\sum_{j,m}j n_{j,m}$ ($E_x=0.5t_x$), tilts the ladder in the real-lattice direction and thus generates a current along $\hat{x}$ (see Methods). After $E_x$ is instantaneously activated, the longitudinal current $J_x(\tau)=2t_x\int^{1}_{-1}\sin(\pi k)n(k,\tau){\rm d}k$ is accessed by measuring the total lattice momentum distribution $n(k,\tau)$ with a band-mapping technique~\cite{zhou2023observation}, where $\tau$ is defined in units of $\hbar/t_x$ and $\hbar$ is the reduced Planck’s constant. The momentum distribution is normalized to the total atom number $N=\int^{1}_{-1}n(k){\rm d}k$, and the lattice momenta $k$ are expressed in units of the real-lattice wavenumber $k_{\rm L}=\pi/d$, with the lattice spacing $d=\SI{380}{nm}$.
The induced transverse Hall polarization $P_y$ is evaluated as a difference in the fractional spin population, with respect to the starting value at $\tau = 0$, namely: $P_y(\tau)=[N_M(\tau)-N_1(\tau)]/N(\tau)-[N_M(0)-N_1(0)]/N(0)$~\cite{zhou2023observation}. The atom number $N_m$ in spin state $m$ is measured by performing an optical Stern-Gerlach detection~\cite{PhysRevLett.105.190401}. To compensate this induced polarization $P_y$, we employ a synthetic gradient realized by controlling the detuning of the Raman coupling (see Methods), equivalent to an external potential term $H_y=-E_y\sum_{j,m}m \,n_{j,m}$, which is activated together with $H_x$.
We determine the Hall voltage $V_{\rm H}$, by adjusting $E_y$ to suitable values such that the time average of the polarization vanishes, i.e. $\langle P_y\rangle=\langle P_y(\tau)\rangle_\tau=0$ in the time interval $\tau\in[1,5]$, for which we have already verified in Ref.~\cite{zhou2023observation} that the Hall response reaches a stationary universal value. We label this fine-tuned field $E_{\rm H}$. It can be immediately put in relation with the Hall voltage by multiplying it by the total number of legs $V_{\rm H}=-(M-1)E_{\rm H}$. Notice that, for the case with two legs, these two quantities coincide. In Fig.~\ref{fig2} we show the dependence of $\langle P_y\rangle$ over $E_y$ for a set of different total atom numbers, used to retrieve the values of the Hall voltage as a function of $N$, as discussed in the following sections.


\vspace{0.5em}
\noindent{\bf Theoretical background -- }
This experimental procedure differs from protocols proposing to measure $\rho_{\rm H}$ either in stationary regimes~\cite{prelovsek1999hall,zotos2000reactive,greschner2019universal,filippone2019vanishing} or to fine-tune $E_{y}$ such to suppress the polarization $P_y$ at every single time~\cite{PhysRevLett.126.030501}, and presents the advantage compared to the proposed protocol of Ref.~\cite{PhysRevLett.126.030501} to work for finite magnetic fields and not just in the limit of $B\to 0$, allowing us to extract a sizeable Hall signal from the experiment. In the Supplementary Information, we show that, regardless of the experimental procedure, for an ideal system -- at zero temperature, infinite and perfectly periodic -- in the single-band metal regime, the polarization $P_y$ and the current $J_x$ are bound by the relation
\begin{equation}\label{eq:univershall}
    \langle P_y\rangle=\left(E_y-2\tan\left(\frac\varphi2\right)\frac{\langle J_x\rangle}N \right)\mathcal I_M\,,
\end{equation}
where only the proportionality constant $\mathcal I_M$ depends on the number of legs $M$ ($\mathcal I_2=1/(2t_y)$ and $\mathcal I_3=\sqrt 2/t_y$, see also Ref.~\cite{greschner2019universal}, which addressed exclusively the $\varphi\rightarrow0$ limit). Additionally, assuming that it is possible to fine-tune $E_y$ to a value $E_{\rm H}$ such that $\langle P_y\rangle=0$, one finds the universal relation 
\begin{equation}\label{eq:rhohexp}
\rho_{\rm H}=\frac{E_{\rm H}}{\langle J_x\rangle}=\frac{2}N\tan\left(\frac\varphi2\right)\,.
\end{equation}
Equations~\eqref{eq:univershall} and~\eqref{eq:rhohexp} feature the total number of atoms $N$, because our experiment can only access global quantities, instead of the current and polarization densities in each tube. We show in the Supplementary Information that such relations reflect the intimate connection between $\rho_{\rm H}$ and the inverse carrier density $1/n$ in each tube. Notice that, as a sign of the universality of the Hall effect, the relation~\eqref{eq:rhohexp} between $E_{\rm H}$ and $J_x$ does not depend on any microscopic detail of the system, apart from the number of carriers $N$ and the magnetic flux $\varphi$. This is not the case for $P_y$, which depends on the microscopic structure through the factor $\mathcal I_M$.

We stress that the presence of strong repulsions in the system is key to stabilize a strongly correlated single-band metal~\cite{greschner2019universal,zhou2023observation,filippone2019vanishing,huang2022topological} and, consequently, the universal scaling Eq.~\eqref{eq:rhohexp}, as we are going  to observe in our experimental setup.


\begin{figure*}[t!]
    \centering
    \includegraphics[width=2\columnwidth]{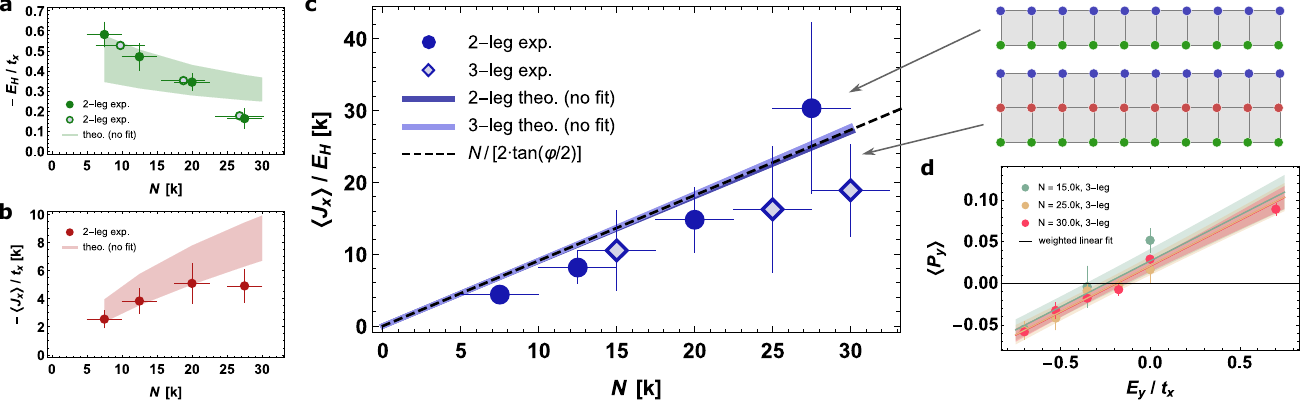}
    \caption{\textbf{Dependence of the inverse Hall resistance on atom number and robustness with respect to ladder geometry.} The data are measured in the single-band regime at $t_y=3.30t_x$ and $U=6.56t_x$. {\textbf{a}}, Hall field $E_{\rm H}$ as a function of the atom number $N$ for a two-leg ladder. The filled circles are obtained from the linear fit of the data in Fig.~\ref{fig2}, and the vertical error bars represent the 95$\%$ confidence interval; the empty circles and their horizontal error bars are also extracted from the data reported in Fig.~\ref{fig2}, when fitting $P_y$ as a function of $N$ for different transverse field values $E_y$.
    {\textbf{b}}, Averaged current $\langle J_x\rangle$ as a function of the atom number $N$ for a two-leg ladder. The vertical error bars denote the standard deviation of $\langle J_x\rangle=\langle J_x(\tau)\rangle_{\tau,{E_y}}$. 
    {\textbf{c}}, Inverse Hall resistance as a function of the atom number $N$ for both two-leg (filled circles) and three-leg (empty diamonds) configurations. The dashed line indicates the universal relation of Eq.~\eqref{eq:rhohexp}. The vertical error bars are obtained with standard uncertainty propagation. The horizontal error bars in {\textbf{a}} (filled circles), {\textbf{b}} and {\textbf{c}} indicate the range of atom number considered for each point. The coloured areas in {\textbf{a}}, {\textbf{b}}, and {\textbf{c}} are the numerical simulations from a mean-field approximation, accounting for the distribution of atom numbers in the tubes and experimental temperature uncertainty $1.5\leq T/t_x\leq 3$.
    {\textbf{d}}, The time-averaged Hall polarization $\langle P_y\rangle$ as a function of the transverse field $E_y$ for three-leg ladders, with different total atom number $N=15{\rm k},\,25{\rm k}$, and $30{\rm k}$. The error bars denote the standard error of the mean, wherein the solid lines in corresponding colours are obtained from a weighted global linear fit with colour shades representing the 95$\%$ confidence bands of the fit.}
    \label{fig3}
\end{figure*}

\vspace{0.5em}
\noindent{\bf Demonstration of the $1/n$ scaling of $\rho_{\rm H}$ -- }
To measure the Hall resistance, we first inspect the behaviour of $\langle P_y\rangle$ in the case of a two-leg ladder, realized upon the nuclear spin states $m_{\rm F}=-1/2$ and $m_{\rm F}=-5/2$. Figure~\ref{fig2} reports the dependence of $\langle P_y\rangle$ as a function of the transverse field $E_y$, for different atom numbers  $N=7.5{\rm k},\,12.5{\rm k},\,20.0{\rm k}$, and $27.5{\rm k}$. Accordingly to the constitutive relation~\eqref{eq:univershall}, $\langle P_y\rangle$ increases linearly with the applied transverse field $E_y$, with a slope that is independent of the atom number $N$. Clearly, the data in Fig.~\ref{fig2} provide a direct verification of this predicted linear dependence (see also Supplementary Fig. S3). On the contrary, the longitudinal current $J_x(\tau)$ does not show a sensitive dependence on $E_y$ (see Supplementary Fig. S2). Thus, we use the current $\langle J_x\rangle=\langle J_x(\tau)\rangle_{\tau,{E_y}}$, averaged over the same time interval $\tau\in[1,5]$ considered for the determination of $P_y$, as well as over different values of transverse field $E_y$.

The linear fit of the experimental data in Fig.~\ref{fig2} allows us to determine the Hall field $E_{\rm H}$ -- or, equivalently, the Hall voltage $V_{\rm H}$ -- corresponding to $\langle P_y\rangle=0$. The dependence of $E_{\rm H}$ with the total atom number $N$ is shown in Fig.~\ref{fig3}a (filled circles). The additional data points in Fig.~\ref{fig3}a (empty circles) are also extracted from the data reported in Fig.~\ref{fig2}, when fitting $P_y$ as a function of $N$ for different transverse field values $E_y$ (see Supplementary Fig. S1). The data show a suppression of $E_{\rm H}$ with the total atom number $N$. Remarkably, the suppression of $E_{\rm H}$ corresponds to the suppression of the longitudinal current per particle $\langle J_x\rangle/N$, which can be inferred from the sub-linear increase of $| \langle J_x \rangle |$ as $N$ is increased, shown in Fig.~\ref{fig3}b. A complete suppression is expected when the single conduction band is entirely filled and thus insulating.

These trends are well reproduced by theoretical simulations relying on a mean-field approximation of the interactions in Eq.~\eqref{eq1}, and reported as coloured areas. As extensively discussed in Ref.~\cite{zhou2023observation}, the mean-field approximation has clear limitations in terms of quantitative accuracy, but it can account for the finite temperatures of the experiment ($T\simeq t_x$), which is much more challenging to address with exact studies. 
Moreover, $P_y$, $J_x$ and $E_{\rm H}$ are all non-universal quantities, which are sensitive to the microscopic details of the experiment, such as the exact distribution of the atoms in the tubes, their temperature, the shape of the drive and confining potential. The width of the coloured areas in Fig.~\ref{fig3} reflects the uncertainty on the temperature of the system, which is the experimental quantity with the highest uncertainty.


However, the fact that these quantities are intimately connected through the relation~\eqref{eq:univershall} puts strong constraints on their ratios, as it was previously shown for the Hall imbalance $\Delta_{\rm H}={P_y}/{J_x}$, which equals the universal value $2t_x\tan(\varphi/2)/t_y$ in the absence of a transverse compensation ($E_y=0$)~\cite{zhou2023observation}.

We now highlight the central result of our work, which emerges as a manifestation of the universality of transverse transport. We consider $\rho_{\rm H}$, extracted from the ratio between the measured values of $E_{\rm H}$ and $\langle J_x\rangle$. Figure~\ref{fig3}c illustrates the linear dependence of the inverse Hall resistance on the atom number. We observe that the experimentally measured $1/\rho_{\rm H}$ exhibits a solid quantitative agreement with the theoretical prediction Eq.~\eqref{eq:rhohexp} (dashed line) for a large range of atom numbers, with no fitting parameters. We also find an almost perfect agreement with $1/\rho_{\rm H}$ as extracted from numerical simulations (coloured area) accounting for different tube average, finite temperatures and confinement. Remarkably, the uncertainty on the temperature, responsible for a major broadening in the numerical prediction of $E_H$ and $\langle J_x\rangle$, has almost no effect on $1/\rho_{\rm H}$, reflecting the universal character of this latter quantity and its robustness against parameter changes.
Our experimental observations testify to the robust and universal connection between the Hall resistance and the inverse carrier density $1/n$ in a strongly correlated single-band metal, stabilized in a two-leg ladder by strong on-rung repulsions.

We further demonstrate the universality of the relation between $\rho_{\rm H}$ and $n$ by modifying the microscopic structure of the system and realizing a three-leg ladder configuration, where the Raman parameters were adjusted to extend the synthetic coupling to the $m_{\rm F}=+3/2$ nuclear spin state (see Methods). The three-leg ladder features a similar trend of $\langle P_y\rangle$ to that of the two-leg case (see Fig.~\ref{fig3}d). Thus we determine $E_{\rm H}$ for the three-leg ladder with the same procedure as for the two-leg ladder and derive the corresponding inverse Hall resistance. Figure~\ref{fig3}c compares the results of two-leg and three-leg ladders for different atom numbers. The measured data for the two configurations follow the same expected universal trend within experimental errors, in agreement with Eq.~\eqref{eq:rhohexp} and numerical simulations.
The consistent relation between $\rho_{\rm H}$ and $n$ in the two- and three-leg ladders confirms the effectiveness of our experimental procedure to measure the Hall resistance of a strongly correlated fermion system.


\vspace{1em}
\noindent\textbf{\large Discussion}

\noindent In this experiment, we have established an efficient protocol to measure the Hall voltage in a fermionic cold-atom system. By employing simultaneous quenches of longitudinal and transverse fields, we have shown distinctive particle-density dependent behaviour of the Hall resistance in a controllable quantum simulator of Hall bars in ladder geometries threaded by a synthetic magnetic flux. These measurements demonstrate the universal relation between the Hall resistance and the number of carrier densities in strongly correlated single-band metals, with remarkable agreement with theoretical predictions.

The ability to measure the Hall voltage and the Hall resistance in clean and highly tunable cold-atom systems opens many exciting possibilities, including the direct comparison with the corresponding measurements in solid-state devices. It paves the way to the investigation of striking features of strongly correlated topological phases of matter, such as the quantization of the transverse resistance (the Hall plateau) with ultracold atoms as well as the role of the interaction anisotropy. Furthermore, interesting perspectives reside in exploring the Hall resistivity of Mott insulating phases at low doping and at the metal to insulator transition~\cite{badoux_change_2016}, as well as measuring other Hall effects such as the thermal Hall one, where the current is induced by a temperature gradient~\cite{Melcer2024}.

\vspace{1em}
\noindent\textbf{\large Methods}

\noindent{\bf Generation of effective magnetic and electric fields -- }
The complex tunneling $t_y e^{i\varphi j}$ along the synthetic dimension in Eq.~\eqref{eq1} is implemented by the coherent coupling between different spin states, via two Raman laser beams propagating with a relative angle. The different periodicity of the Raman coupling with respect to the lattice spacing $d$, gives rise to a non-zero Peierls phase $\varphi=(\Delta\vec{k}\cdot\hat{x})d$ with $\Delta\vec{k}$ denoting the wavevector of the coupling field, leading to a synthetic magnetic flux $\varphi=0.32\pi$ piercing the ladder plaquettes.

The effective electric field $E_x$ along the longitudinal direction is realized by shining a far-off-resonance red-detuned laser beam operating at $\SI{1112}{nm}$. This laser is aligned in the way that the atomic cloud center is located at the maximum slope of the Gaussian beam profile, to ensure the induced light-shift gradient, which stems from the intensity gradient provided by the Gaussian beam, has sufficient linearity. In particular, the beam waist is $\SI{85}{\mu m}$ and the variation of the expected Bloch frequency among central 40 lattice sites (estimated region of the atomic sample) is less than 3$\%$.

The effective electric field $E_y$ along the synthetic dimension is instead implemented by abruptly adding a small Raman laser detuning ($\SI{-120}{Hz}\sim\SI{120}{Hz}$) after adiabatically loading the ground state of the system. To relate such a detuning to an energy difference along the synthetic dimension, consider the Hamiltonian describing the Raman coupling for a two-leg ladder in the rotating frame
\begin{equation}
H_{{\rm R}} =
\begin{pmatrix}
0   & t_y \\
t_y & -E_y \\
\end{pmatrix}\,.
\end{equation}
One can clearly see that the presence of a detuning on the diagonal terms directly translates into an energy shift for different spin states, namely an energy variation in the synthetic dimension. The Hall voltage $V_{\rm H}$ is then directly proportional to the fine-tuned Hall field $E_{\rm H}$, which is expressed as $V_{\rm H}=-(M-1)E_{\rm H}$ by accounting for the total number of legs $M$.

\vspace{0.5em}
\noindent{\bf Initial state preparation -- }
The $^{173}$Yb atoms are initially polarized in the $\lvert F=\sfrac52,\,m_{\rm F}=-\sfrac52\rangle$ hyperfine component of the electronic ground state with a typical temperature of $0.2 \sim 0.25 T_{\rm F}$, where $T_{\rm F}$ is the Fermi temperature. The atomic cloud is first loaded into the vertical optical lattice along the gravitational direction within $\SI{150}{ms}$ using an exponential intensity ramp. The optical dipole trap, where the atoms are initially confined, is subsequently switched off in $\SI{1}{s}$. Two horizontal optical lattices are then ramped up in the same way as the vertical lattice. The vertical lattice depth is set to $15 E_{\rm r}$, where $E_{\rm r}=h^2/8m d^2$ is the recoil energy, $h$ is the Planck constant, $m$ is the atomic mass, and $d$ is the lattice spacing. Two horizontal lattice depths are set to $15 E_{\rm r}$ and $4 E_{\rm r}$, respectively, between which the shallow one with lattice depth of $4 E_{\rm r}$ is along the $\hat{x}$-direction. Thus, the tunneling rate $t_x/\hbar$ ($2\pi\times \SI{171}{Hz}$) along the fermionic tubes is much larger than the radial tunneling rates ($2\pi\times \SI{12.96}{Hz}$), which ensures the dynamics is only allowed in the shallow lattice along longitudinal direction $\hat{x}$. These independent 1D fermionic tubes are characterized by an axial harmonic confinement with a trapping frequency of $2\pi\times\SI{43}{Hz}$, which originates from the Gaussian intensity profiles of the 2D red-detuned lattice beams.

After the lattice loading procedure, we employ an adiabatic preparation sequence to slowly activate the tunnel coupling between the legs and produce the equilibrium state of a $M$-leg ladder, where the nuclear spins act as different sites along a synthetic dimension $\hat{y}$ and whose plaquettes are threaded by a synthetic magnetic flux $\varphi$. The Raman laser beams are switched on with an initial detuning $\delta_{\rm i}=\SI{-10}{kHz}$ and perform an exponential frequency sweep of the form
\begin{equation}
\label{delta_t}
\delta(t)=\delta_{\rm i}+(\delta_{\rm f}-\delta_{\rm i}) \left(\frac{1-e^{-t/T_{\rm{tau}}}}{1-e^{-T_{\rm{adiab}}/T_{\rm{tau}}}}\right)\,,
\end{equation}
where $\delta_{\rm f}$ is chosen to resonantly couple the two nuclear spin states $\lvert m_{\rm F}=-1/2\rangle$ and $\lvert m_{\rm F}=-5/2\rangle$, with the ramp duration $T_{\rm{adiab}}=\SI{30}{ms}$ and $T_{\rm{tau}}=\SI{14}{ms}$. The adiabaticity of the whole process is verified experimentally by reversing the whole procedure to recover a spin-polarized Fermi gas.

The Raman beams couple up to three nuclear spin states $m_{\rm F}=-\sfrac52,\,-\sfrac12$ and $+\sfrac32$ via $\sigma^{+}/\sigma^{-}$ Raman transitions. The atoms are subjected to a $B_0=\SI{153}{Gauss}$ (real) magnetic field along the vertical direction, generating a linear Zeeman splitting $\Delta_Z=\SI{31.6(7)}{kHz}$ between adjacent nuclear spin components. The frequency difference between the two Raman beams is set to $2\Delta_Z$ and we exploit the polarization-dependent Raman-induced light shifts to switch between the two-leg and the three-leg ladder case~\cite{science.aaa8736}. For the two leg case, we implement a horizontal polarization for both Raman laser beams, relative to the vertical quantization axis defined by the magnetic field, i.e. in the spherical basis, $\hat{\epsilon}_{2L}=(\hat{\epsilon}_{+}+\hat{\epsilon}_{-})/\sqrt{2}$. Having a purely horizontal polarization offsets the $\lvert m_{\rm F}=+3/2\rangle$ spin state and effectively isolates it from the dynamics, being its population at most a few percent. To instead realize a three leg ladder, all three states need to be resonantly coupled. This condition can be fulfilled by implementing a uniform polarization for both Raman beams, namely $\hat{\epsilon}_{3L}=(\hat{\epsilon}_{+}+\hat{\epsilon}_{-}+\hat{\epsilon}_{\pi})/\sqrt{3}$. Under this condition the light shifts are approximately the same for the three states, which are all substantially populated throughout the experiment. The Raman coupling strength has a slight asymmetry for different nuclear spin states in the three-leg configuration owing to the spin-dependent Clebsch-Gordan relations. However, because the experiment is performed in the universal single-band regime, the specific value of $t_y$ has no significant impact on the experimental results.

\vspace{0.5em}
\noindent{\bf Theoretical methods and comparison to experimental data -- }
The universal expression Eq.~\eqref{eq:univershall} for the Hall response is obtained by performing the perturbative expansion of the polarization $P_y$ to leading order in the longitudinal hopping  $t_x$ and in the transverse field  $E_y$, and by assuming that the strong interactions ($U=6.56t_x$) stabilize a single band metal (see Supplementary Fig. S4)~\cite{greschner2019universal,filippone2019vanishing}. 

The details of this lengthy, but straightforward calculation are given in the Supplementary Information, where we also discuss the mean-field approximation and its comparison with the experimental data.

\vspace{1em}
\noindent\textbf{\large Data availability}

\noindent All the experimental and theoretical data presented in the figures of the main article and Supplementary Information are available for download from an open repository~\cite{Zenodo}.

\vspace{1em}
\noindent\textbf{\large Code availability}

\noindent The numerical codes used in this study are publicly accessible at \url{https://github.com/michelefilippone/Measuring_Hall_voltage_and_Hall_resistance_in_an_atom-based_quantum_simulator_OPEN_SOURCE}.


\vspace{1em}
\noindent\textbf{\large References}

\bibliography{apssamp}

\vspace{1em}
\noindent\textbf{\large Acknowledgements}

\noindent We gratefully acknowledge J. Mellado Mu\~noz for discussions and critical reading of the manuscript. For the experimental activity we acknowledge financial support by PNRR MUR project PE0000023-NQSTI financed by the European Union - Next Generation EU and by the  Horizon Europe programme HORIZON-CL4-2022-QUANTUM-02-SGA via the project 101113690 (PASQuanS2.1). This work is supported in part by the Swiss National Science Foundation under grant 200020\_219400. M.F. acknowledges support from EPiQ ANR-22-PETQ-0007 part of Plan France 2030. C.R. acknowledges support from ANR through Grant No. ANR-22-CE30-0022-01.

\vspace{1em}
\noindent\textbf{\large Author contributions}

\noindent L. F., J. C., G. C., M. F., T.-W. Z., and T. G. conceived the experiments. T.-W. Z., T. B., G. M., and J. P. carried out the experimental work. T.-W. Z., T. B., and G. M. analyzed the experimental results. C. R., M. F., and T. G. performed theoretical work. All authors contributed extensively to the discussion of the results and to the writing of the manuscript. 

\vspace{1em}
\noindent\textbf{\large Competing interests}

\noindent The authors declare no competing interests.


\newpage
\clearpage

\setcounter{page}{1}

\renewcommand{\thefigure}{\arabic{figure}}
\renewcommand{\figurename}{SUPP.\ FIG.}
 \setcounter{figure}{0}
\renewcommand{\theequation}{S.\arabic{equation}}
 \setcounter{equation}{0}
\renewcommand{\thesection}{S.\Roman{section}}
\setcounter{section}{0}
\setcounter{secnumdepth}{3}
\renewcommand{\thetable}{S\arabic{table}}
 \setcounter{table}{0}

\onecolumngrid

\begin{center}
{\bf \large Supplementary Information for\\
\vspace{3mm}
``Measuring Hall voltage and Hall resistance in an atom-based quantum simulator''}\\
\vspace{3mm}
T.-W. Zhou, T. Beller, G. Masini, J. Parravicini, G. Cappellini,\\
C. Repellin, T. Giamarchi, J. Catani, M. Filippone, L. Fallani
\vspace{3mm}
\end{center}

\onecolumngrid

\section{Data analysis}

As a sanity check of the Hall voltage measurement, we reorganize the same data in Fig.~\ref{fig2} but in a different way. Instead of grouping the data by atom number $N$, the data shown in Supplementary Fig.~\ref{figE1} are grouped by transverse field $E_y$. Here, only three out of six groups of data, which exhibit a sign change of $\langle P_y\rangle$, are kept since the target is still to find the condition where the time-averaged Hall polarization $\langle P_y\rangle=0$. Thus, the three additional Hall voltage data points (empty circles in Fig.~\ref{fig3}a) are obtained from the linear fit in Supplementary Fig.~\ref{figE1}. The outcome indicates that the whole Hall voltage measurement is remarkably robust.

\begin{figure}[h!]
    \centering
    \includegraphics[width=.55\columnwidth]{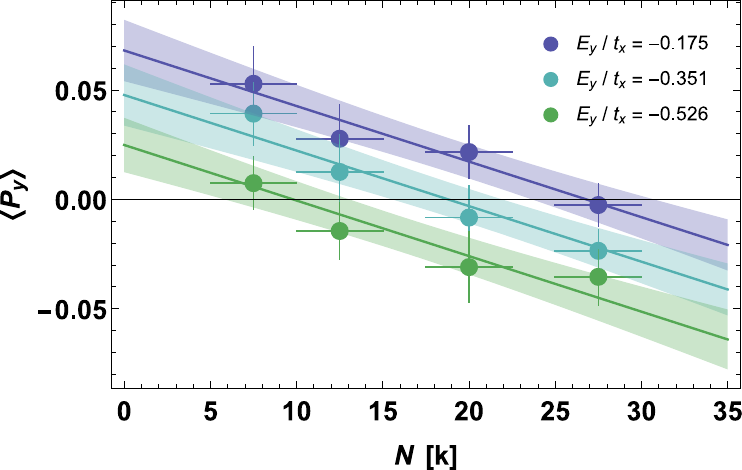}
    \caption{Time-averaged Hall polarization $\langle P_y\rangle$ for two-leg ladders, measured at $t_y=3.30t_x$ and $U=6.56t_x$, as a function of the atom number $N$ with different transverse field $E_y$. The horizontal and vertical error bars depict the measurement interval and the standard error of the mean obtained with a statistical Bootstrap method, respectively. The solid lines in corresponding colours are obtained from a weighted global linear fit based on the error bars of the experimental data, while colour shades represent the 95$\%$ confidence bands of the fit.}
    \label{figE1}
\end{figure}

Besides, in order to construct a fitting model which incorporates all the data points for different transverse field $E_y$, as well as different atom number $N$, the solid lines in Fig.~\ref{fig2}, Fig.~\ref{fig3}d and Supplementary Fig.~\ref{figE1} are obtained by performing a weighted global linear fit, wherein the vertical error bars of the data points are used as weights in the linear fit function, and the shared parameter amongst the different datasets, grouped by $N$ (Fig.~\ref{fig2} and Fig.~\ref{fig3}d) or $E_y$ (Supplementary Fig.~\ref{figE1}), is the slope of the fitting curves.

\begin{figure}[h!]
    \centering
    \includegraphics[width=.625\columnwidth]{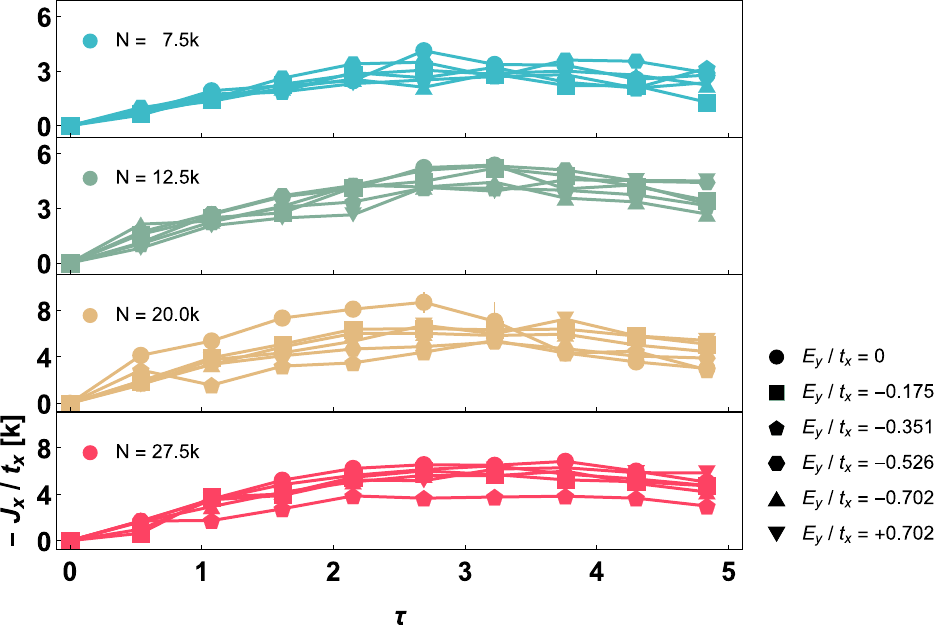}
    \caption{Time evolution of the longitudinal current $J_x$ for two-leg ladders, measured at $t_y=3.30t_x$ and $U=6.56t_x$, with different atom number $N$ and transverse field $E_y$. The error bars represent the standard error of the mean and are obtained with a statistical Bootstrap method.}
    \label{figE2}
\end{figure}

As mentioned in the main text, the current $\langle J_x\rangle=\langle J_x(\tau)\rangle_{\tau,{E_y}}$ is obtained by averaging over the time interval $\tau\in[1,5]$ and different values of $E_y$. In Supplementary Fig.~\ref{figE2} we show the time evolution of the longitudinal current $J_x$ for two-leg ladders with different atom number $N$ and transverse field $E_y$, it is plain to see that $J_x(\tau)$ does not exhibit a sensitive dependence on $E_y$.

\begin{figure}[h!]
    \centering
    \includegraphics[width=.55\columnwidth]{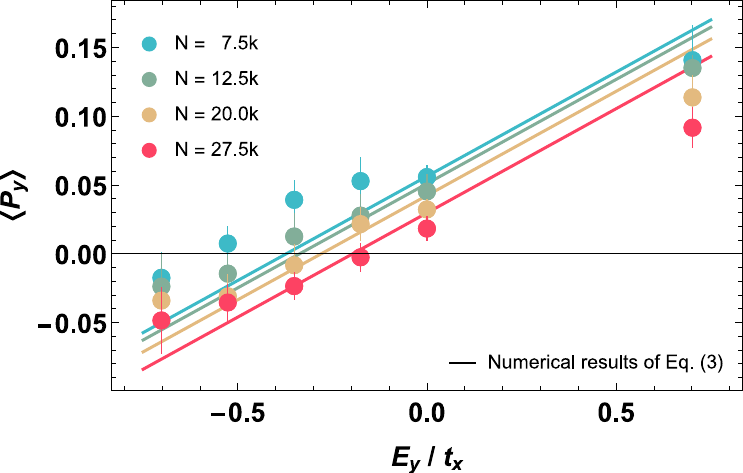}
    \caption{The time-averaged Hall polarization $\langle P_y\rangle$ for two-leg ladders, measured at $t_y=3.30t_x$ and $U=6.56t_x$, is shown as a function of the transverse field $E_y$ with different atom number $N$, the data points are the same as reported in Fig.~\ref{fig2}. Instead, the solid lines in corresponding colours depict the numerical results of Eq.~\eqref{eq:univershall} based on the experimental $\langle J_x\rangle$.}
    \label{figE3}
\end{figure}

Based on the experimentally measured $\langle J_x\rangle$, we also make a calculation of $\langle P_y\rangle$ according to Eq.~\eqref{eq:univershall}. The results are shown as the solid lines in Supplementary Fig.~\ref{figE3}, together with the experimentally measured $\langle P_y\rangle$. As we can see from Supplementary Fig.~\ref{figE3}, the relation~\eqref{eq:univershall} works remarkably well at $E_y=0$, revealing that the universal Hall imbalance $\Delta_{\rm H}={P_y}/{J_x}$ equals the value $=2t_x\tan(\varphi/2)/t_y$ in the single-band metal regime. The relation~\eqref{eq:univershall} also captures the overall linear dependence of the polarization $P_y$ on the transverse field $E_y\neq0$. The deviations can be ascribed to the fact that all these quantities, when considered individually, strongly depend on the microscopic details of the experimental protocol, while their ratios display a more robust behaviour, as we have shown in the main text.


\section{Theoretical analysis}

\begin{figure}[b!]
    \centering
    \includegraphics[width=.7\columnwidth]{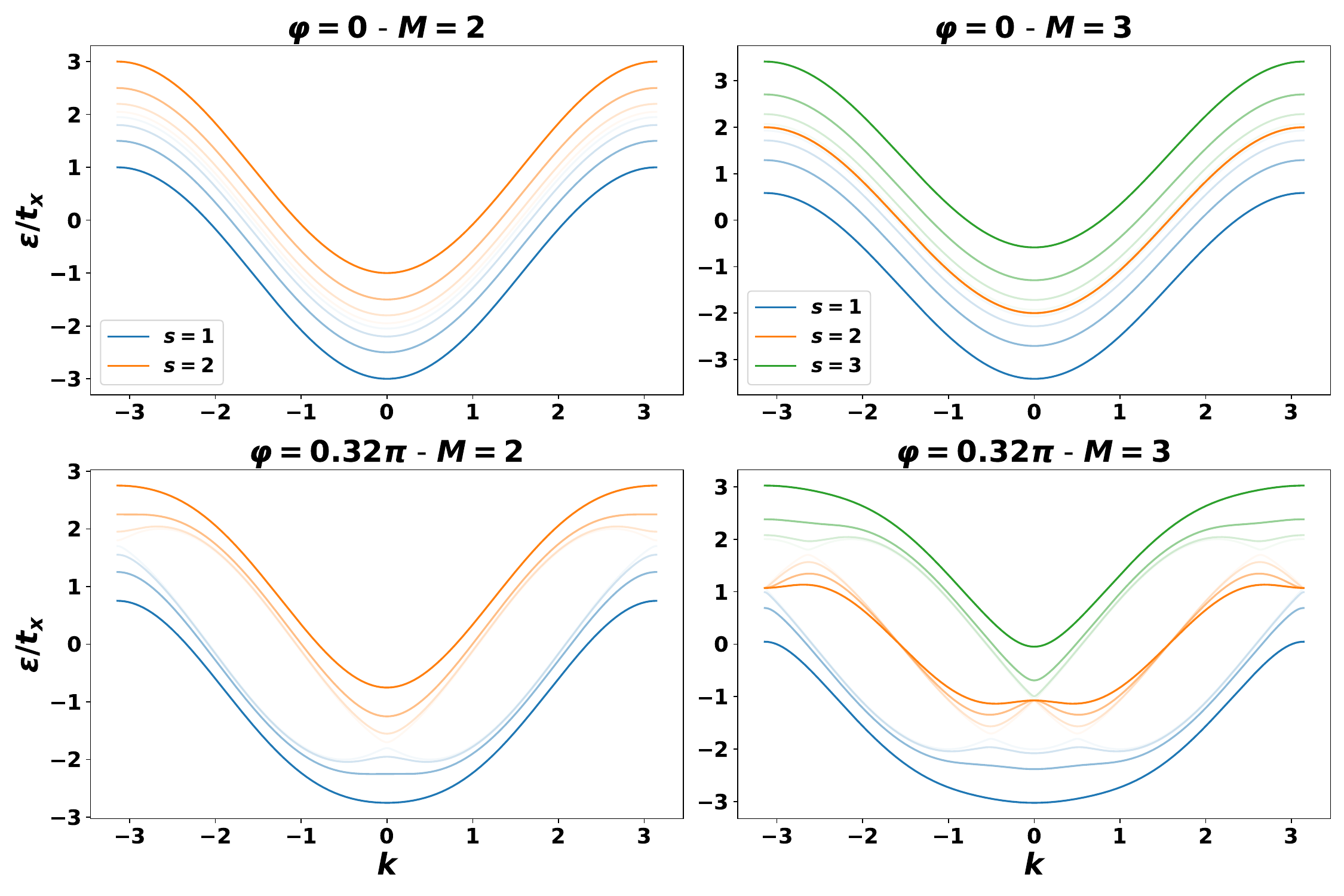}
    \caption{Single-particle band spectrum for $M=2$ and $M=3$ legs (left and right respectively) as function of the quasi-momentum $k$ along the $\hat x$-direction. On the top we consider the case without flux ($\varphi=0$) while on the bottom we consider the experimental flux ($\varphi=0.32\pi$). The different colors correspond to different band indexes $s$, while lines from light to dark corresponds to increasing values of the transverse tunneling amplitude $t_y/t_x=0.05,\,0.2,\,0.5,\,1$.}
    \label{figS1}
\end{figure}

In this section we derive the universal relation~\eqref{eq:univershall} in the main text for ideal single-band metals which are translational invariant along the $\hat x$-direction. We start with the expression of the polarization density $p_y$ 
\begin{align}\label{eq:py}
    p_y&=\frac{c}{LM}\sum_{j,m=1}^M(m_0-m)\langle a^\dagger_{j,m}a^{\phantom\dagger}_{j,m}\rangle\,, 
\end{align}
where we have introduced the number of sites $L$ along the $\hat x$-direction and $m_0=(M+1)/2$ is the center coordinate of the ladder system along the synthetic dimension. The constant $c$ is introduced for consistency with the global polarization defined in the main text,  $c=$2 or 1 depending on whether $M=2$ or 3  respectively. To exploit translation invariance along the longitudinal direction $\hat x$, we perform the  gauge transformation  $a_{j,m}\rightarrow a_{j,m}e^{-i\varphi j(m-m_0)}$ and assume  periodic boundary conditions, leading to 
\begin{equation}
H=-t_x\sum_{j,m=1}^{M}\Big[e^{-i\varphi(m-m_0)}a^\dagger_{j,m}a^{\phantom\dagger}_{j+1,m}+\mbox{h.c.}\Big]-t_y\sum_{j,m=1}^{M-1}\Big[a^{\dagger}_{j,m}a^{\phantom\dagger}_{j,m+1}+\mbox{h.c.}\Big]+E_y\sum_{j,m=1}^{M}(m-m_0)n_{j,m}\,.
\end{equation}
We omit the interaction term of Eq.~\eqref{eq1} in the main text for the moment. We have also added a term proportional to the total atom number -- $E_ym_0N$ -- to use consistent notations for the synthetic dimension $m$. For the following discussion, all the boundary effects associated to the gauge transformation, discussed in Refs.~\cite{PhysRevB.97.201408,PhysRevB.99.035150}, can be safely neglected. We switch to $k$-space, where the Hamiltonian reads
\begin{equation}\label{eq:ham}
H=-t_y\sum_{k,m=1}^{M-1}\Big[c^{\dagger}_{k,m}c^{\phantom\dagger}_{k,m+1}+\mbox{h.c.}\Big]+\sum_{k,m=1}^{M}\Big[-2t_x\cos\left(k-(m-m_0)\varphi\right)+E_y(m-m_0)\Big]n_{k,m}\,,
\end{equation}
where $n_{k,m}=a^\dagger_{k,m}a^{\phantom \dagger}_{k,m}$. 
This Hamiltonian is diagonalized by applying the unitary transformation $a_{k,m}=\sum_s u_{m,s}(k,\varphi)c_{k,s}$, where $s$ is the band index and the coherence factors $u_{m,s}$ define the unitary diagonalizing the non-interacting part of Eq.~\eqref{eq1} in the main text. We will omit their $k$ dependence for compactness now on. Typical plots of the single-particle band spectrum as function of the quasi-momentum $k$ are shown in Supplementary Fig.~\ref{figS1}. In particular, in the $\varphi,E_y\rightarrow0$ limit, the non-interacting spectrum decomposes in a longitudinal and transverse contribution $\varepsilon^0_{k,s}=\varepsilon^0_k+\varepsilon^0_s$, with $\varepsilon^0_k=-2t_x\cos(k)$ and $\varepsilon^0_s=-2t_y\cos[\pi s/(M+1)]$ with $s\in[1,M]$. Equation~\eqref{eq:py} then becomes 
\begin{equation}\label{eq:pyrotared}
    p_y=\frac{c}{LM}\sum_{k,m,s,s'}(m_0-m)u^*_{m,s}u^{\phantom*}_{m,s'}\langle c^\dagger_{k,s}c^{\phantom\dagger}_{k,s'}\rangle\,.
\end{equation}
For convenience, we also add the corresponding expression for the current density 
\begin{equation}
    j_x=\frac{2t_x}{LM}\sum_{k,m,s,s'}\sin(k-\varphi(m-m_0))u^*_{m,s}u^{\phantom*}_{m,s'}\langle c^\dagger_{k,s}c^{\phantom\dagger}_{k,s'}\rangle\,.
\end{equation}
We will consider two limits: the one of weak magnetic fluxes ($M\varphi\ll2\pi$) and the one of large intra-leg coupling ($t_y\gg t_x$).

\subsection{Weak magnetic fluxes -- $M\varphi\ll2\pi$}\label{sec:smallphi}
We recur to standard perturbation theory. Expanding Eq.~\eqref{eq:ham} to leading order in $\varphi$ and $E_y$, we can cast the Hamiltonian in the form  $ H= H_0+ V\,,$ with 
\begin{equation}
V=\sum_{k,m,s,s'}(m-m_0)[-2t_x\sin(k)\varphi+E_y]\times u^{0}_{m,s}u^0_{m,s'}c^\dagger_{k,s}c^{\phantom \dagger}_{k,s'}\,.
\end{equation}
Notice again that the perturbation $V$ is diagonal in $k$-space and the coherence factors $u^0_{m,s}=\sqrt2\sin[\pi s\,m/(M+1))]/\sqrt{M+1}$, with $s\in[1,M]$ are also $k$- and $\varphi$-independent. Applying standard perturbation theory, one finds the corrections to the coherence factors $u_{m,s}=u_{m,s}^0+u_{m,s}^1$, where 
\begin{equation}
\begin{split}
u^1_{m,s}=&[-2t_x\sin(k)\varphi+E_y]\times\sum_{m',s'\neq s}\frac{(m'-m_0)u^{0}_{m',s'}u^{0}_{m',s}}{\varepsilon_s^0-\varepsilon_{s'}^0}u^0_{m,s'}\,\,,
\end{split}
\end{equation}
and where $\varepsilon^0_s$ is the transverse component of the non-interacting spectrum reported before Eq.~\eqref{eq:pyrotared}. Inserting this last expression in Eq.~\eqref{eq:pyrotared}, we find   
\begin{equation}\label{eq:pysmallflux}
p_y=\frac c{LM}\sum_{k,s,s'}[2t_x\sin(k)\varphi-E_y]\langle c^\dagger_{k,s}c^{\phantom\dagger}_{k,s'}\rangle\left[I^{(M)}_{s,s'}+I^{(M)}_{s',s}\right]\,,
\end{equation}
with 
\begin{equation}\label{eq:Is}
\begin{split}
I^{(M)}_{s,s'}=\sum_{\stackrel{m,m'}{s''\neq s}}&\frac{(m-m_0)(m'-m_0)}{\varepsilon^0_{s}-\varepsilon^0_{s''}}u^0_{m,s}u^0_{m',s''}u^0_{m',s'}u^0_{m,s''}\,.
\end{split}
\end{equation}
If only the lowest band ($s=1$) contributes to transport, that is only $\langle c^\dagger_{k,1}c^{\phantom\dagger}_{k,1}\rangle \neq0$, one finds
\begin{equation}\label{eq:pysmallvarphiuniversal}
p_y=c\Big(\varphi j_x-E_y \,n\Big)2I^{(M)}_{1,1}\,,
\end{equation}
where $n$ is the total particle density and 
\begin{equation}
j_x=\frac{2t_x}{L M}\sum_{k}\sin(k)\langle n_{k,s=1} \rangle
\end{equation}
is the current density flowing in the system along the $\hat x$ direction in the $\varphi\rightarrow0$ limit. As a numerical application, for the case $M=2,\,3$ one readily finds
\begin{align}
I^{(2)}_{1,1}&=-\frac1{8t_y}\,, & I^{(3)}_{1,1}&=-\frac1{2\sqrt2t_y}\,. 
\end{align}
For instance, for $E_y=0$, one readily finds the universal relation for the Hall imbalance $\Delta^{(M)}_{\rm H,\varphi\rightarrow0}=\lim_{\varphi\rightarrow0}\frac{p_y}{ \varphi j_x}$ at weak fluxes~\cite{PhysRevLett.122.083402} 
\begin{align}\label{eq:deltavarphi}
\Delta_{\rm H,\varphi\rightarrow0}^{(2)}&=-\frac1{2t_y}\,,& \Delta_{\rm H,\varphi\rightarrow0}^{(3)}&=-\frac1{\sqrt2t_y}\,.
\end{align}
Notice that there is a global difference of sign and a difference of a factor 2 with $\Delta_{\rm H,\varphi\rightarrow0}^{(3)}=\sqrt2/t_y$ of Ref.~\cite{PhysRevLett.122.083402}. The results are however consistent because there, the polarization is always defined as $P_y=2\sum_{j,m}(m-m_0)\langle n_{j,m}\rangle$, regardless of the number of legs $M$ and the opposite convention for the flux $\varphi\rightarrow-\varphi$ is taken.

\subsection{Large intra-leg coupling --  $t_y\gg t_x$}\label{sec:smalltx}
An analogous calculation can be performed in the $t_y\gg t_x$ limit. In this case the perturbation is provided by the second term in Eq.~\eqref{eq:ham}, namely 
\begin{equation}
V=\sum_{k,m=1}^{M}\Big[-2t_x\cos\left(k-(m-m_0)\varphi\right)+E_y(m-m_0)\Big]n_{k,m}\,.
\end{equation}
This perturbation leads to a modified version of Eq.~\eqref{eq:pysmallflux} 
\begin{equation}
\begin{split}
p_y=\frac c {LM}\sum_{k,s,s'}\langle c^\dagger_{k,s}c^{\phantom\dagger}_{k,s'}\rangle&\left\{2t_x\left[L^{(M)}_{s,s'}(k)+L^{(M)(k)}_{s',s}\right]\right.\left.-E_y\left[I^{(M)}_{s,s'}+I^{(M)}_{s',s}\right]\right\}\,,
\end{split}
\end{equation}
where we have introduced the additional quantity 
\begin{equation}\label{eq:L}
\begin{split}
L^{(M)}_{s,s'}(k)&=\sum_{m,m',s''\neq s}\,\frac{\cos(k-\varphi(m-m_0))(m'-m_0)}{\varepsilon^0_s-\varepsilon^0_{s''}}u^0_{m,s}u^0_{m',s''}u^0_{m',s'}u^0_{m,s''}\,.
\end{split}
\end{equation}
Also in this case, we restrict to the case where only the lowest band  ($s=1$) contributes to transport, that is only $\langle c^\dagger_{k,1}c^{\phantom\dagger}_{k,1}\rangle \neq0$. The coherence factors $u_{m,s}$ are either odd or even functions around $m_0$. As a consequence, splitting the cosinus in Eq.~\eqref{eq:L}, the contribution proportional to \begin{equation}
\left(\sum_{m}\cos(\varphi(m-m_0))u^0_{m,0}u^0_{m,s''}\right)\cdot\left(\sum_{m'}(m'-m_0)u^0_{m',0}u^0_{m',s''}\right)=0\,.
\end{equation}
The reason is that it is the product of two sums convolving the same odd or even function $u_{m,0}u_{m,s''}$ with either an even or odd function around $m_0$ respectively, that is $\cos(\varphi(m-m_0))$ or $(m-m_0)$. Thus, the  polarization reads
\begin{equation}\label{eq:pylargety}
    p_y=2c\Big(j_xG^{(M)}_{0,0}-E_y \,n\,I^{(M)}_{0,0}\Big)\,,
\end{equation}
with
\begin{equation}
G^{(M)}_{0,0}=\frac1{ \sum_{m=1}^{M}\cos(\varphi(m-m_0))(u^0_{m,1})^2}\sum_{m,m',s''}\frac{\sin(\varphi(m-m_0))(m'-m_0)}{\varepsilon_1-\varepsilon_{s''}}u^0_{m,1}u^0_{m,s''}u^0_{m',1}u^0_{m',s''}\,,
\end{equation}
where the current reads
\begin{equation}
    j_x=2t_x\left(\frac1{LM}\sum_k\sin(k)\langle c^\dagger_{k,1}c^{\phantom\dagger}_{k,1}\rangle\right)\times\\\left( \sum_{m=1}^{M}\cos(\varphi(m-m_0))(u^0_{m,1})^2\right)\,.
\end{equation}
Notice that the small field limit $\varphi\rightarrow 0$ maps Eq.~\eqref{eq:pylargety} onto the universal expression~\eqref{eq:pysmallvarphiuniversal}. For $M=2$ and 3, one finds
\begin{align}
    G_{0,0}^{(2)}&=-\frac1{4t_y}\tan\left(\frac\varphi2\right)=2\tan\left(\frac\varphi2\right)I_{0,0}^{(2)}\,, & G_{0,0}^{(3)}&=-\frac{1}{\sqrt2 t_y}\tan\left(\frac\varphi2\right)=2\tan\left(\frac\varphi2\right)I_{0,0}^{(3)}\,,
\end{align}
which allows to cast Eq.~\eqref{eq:pylargety} in a similar form as Eq.~\eqref{eq:pysmallflux}, found for small fields, namely
\begin{equation}\label{eq:pylargetycompact}
   p_y^{(2,3)}=2c\left[2\tan\left(\frac\varphi2\right)j_x-E_y \,n\right]I_{0,0}^{(2,3)}\,.
\end{equation}
Sending $E_y=0$ in Eq.~\eqref{eq:pylargetycompact}, one finds the universal value probed in Ref.~\cite{Zhou2023} (we multiply by $2t_x$ as the current there was normalized by that factor)
\begin{equation}\label{eq:deltaty}
\Delta_{\rm H}^{(2)}=\frac{2t_xp_y}{j_x}=\frac{2t_x}{t_y}\tan\left(\frac\varphi2\right)\,.
\end{equation}

Notice that the above arguments leading to Eqs.~\eqref{eq:deltavarphi} and~\eqref{eq:deltaty}, which are intimately related to the relations~\eqref{eq:univershall} and~\eqref{eq:rhohexp} discussed in the main text, they strictly rely on the assumption that the system is in a metallic state, where only the lowest band is occupied. As long as interactions do not break this assumption -- as a matter of fact, they may even enforce it, as it is the case of the SU($M$) interaction studied here~\cite{PhysRevLett.122.083402,Zhou2023,PhysRevLett.123.086803} -- their detailed form should not modify these results. The effects on the Hall response of intra-leg interactions were addressed in Ref.~\cite{PhysRevLett.123.086803}, but in a different context. A systematic study of the effects of different kinds of interactions on the Hall response of fermionic ladders deserves further investigation.

\section{Conversion to experimental quantities}

Equation~\eqref{eq:univershall} in the main text is obtained by making the connection between the current and polarization densities -- $j_x$ and $p_y$ --  with the experimental quantities -- $J_x$ and $P_y$. Let us consider the total number of atoms $N$ in the experiment. To derive the conversion we start by considering the expression of the total number of atoms
\begin{equation}
    N=\int_{-1}^1n(k)dk  =\sum_{m,k,\rm tubes }f_{\rm tube}(k,m)\,,
\end{equation}
where $f_{\rm tube}(k,m)$ is the momentum distribution of the spin species $m$ in a specific tube. Considering standard transformations, we get, for a uniform system of size $L$, 
\begin{equation}
\begin{split}
    N&=\sum_{m,k,\rm tubes}f_{\rm tube}(k,m)=\frac {L}{2}\int_{-1}^1 dk\,\sum_{m,\rm tubes} f_{\rm tube}(\pi k,m)\,.
\end{split}
\end{equation}
We thus get the equality 
\begin{equation}
    n(k)=\frac {L}{2} \sum_{m,\rm tubes}f_{\rm tube}(\pi k,m)\,.
\end{equation}
If we consider the definition of the experimental current, we can recast it in the following form
\begin{equation}
\begin{split}
    J_x&=2t_x\int_{-1}^1 dk\,\sin(\pi k)\,n(k)=2t_x\sum_{k,m,\rm tubes}\sin(k)f_{\rm tube}(k,m)\,.
\end{split}
\end{equation}
By introducing the current density per tube in the original gauge
\begin{equation}
\begin{split}
j_{x,\rm tube}
&=2t_x\frac1{LM}\sum_{k,m}\sin(k)f_{\rm tube}(k,m)\,,
\end{split}
\end{equation}
and the analogous formula for the polarization, we find the connection between the theoretical  densities and the experimental quantities
\begin{align}
P_y&= \frac{ L M}{N}\sum_{\rm tubes}p_{y,\rm tube}\,,&J_x&=L M\sum_{\rm tubes}j_{x,\rm tube}\,.
\end{align}
Notice that the current normalization has been modified with respect to Ref.~\cite{Zhou2023}.  
The experimental equivalent of Eq.~\eqref{eq:pylargetycompact} becomes Eq.~\eqref{eq:univershall} given in the main text
\begin{align} P_y^{(2)}&=\frac1{2t_y}\left[-2\tan\left(\frac\varphi2\right)\frac{ J_x}N+E_y\right]\,,&  P_y^{(3)}&=\frac{\sqrt2}{t_y}\left[-2\tan\left(\frac\varphi2\right)\frac{J_x}N+E_y\right]\,.
\end{align}
Notice that such relation does not change if averages over the time evolution are taken.  Finding the electric field $E_y$ such that $\langle P_y\rangle =0$ leads to the Hall conductance in the single-band limit given by Eq.~\eqref{eq:rhohexp} in the main text.

\section{Mean-field approximation and Numerical simulations}

It is challenging to simulate the experimental conditions exactly as the experiment features thousands of tubes with up to $40$ interacting atoms, roughly $200\times M$ accessible sites and temperatures $T$ of the order of the longitudinal hopping $t_x$. We thus rely on a mean-field approximation of the repulsive interactions in Eq.~\eqref{eq1}, while none of the aspects of the protocol to measure the Hall voltage as well as the universal behavior of the Hall resistance are dependent on this mean field approximation. The clear limitations of this approximation were extensively discussed in Ref.~\cite{Zhou2023}, but they are convenient to give a semi-quantitative account of the effects of finite temperatures and other parameters in the experimental observations. We give here a short account to justify why the mean-field analysis predicts an increase value of the transverse hopping $t_y$.

We focus on the case with two legs and $E_y=0$. The mean-field decoupling of Eq.~\eqref{eq1} reads
\begin{equation}
\begin{split}
U\sum_{j=1}^L&n_{j,2}n_{j,1}\simeq U\sum_j\Big[n_{j,2}\av{n_{j,1}}+n_{j,1}\av{n_{j,2}}-a^{\dagger}_{j,2}a_{j,1}\av{a^\dagger_{j,1}a_{j,2}}-a^{\dagger}_{j,1}a_{j,2}\av{a^\dagger_{j,2}a_{j,1}}\Big]\,,
\end{split}
\end{equation}
where we have discarded the constant contributions to the  total energy. If we assume equilibrium (no current flowing in the system) and that interactions do not lead to spontaneous breaking of translational invariance, averages do not depend on the lattice rung and leg labels $(j,m)$. We can thus replace the average local occupations with the density $\av{n_{j,m}}=n$. This substitution  leads to the standard Hartree renormalization of the chemical potential. 

Additionally, we find that interactions also lead to a renormalization of the transverse hopping $t_y$ 
\begin{align}\label{eq:typ}
t_y&\longrightarrow t^*_y=t_y+U\,\Omega_{\varphi,U,\rho}\,,
\end{align}
with $\Omega_{\varphi,U,\rho}=\sum_j\langle a^\dagger_{j,0}a_{j,1}\rangle/L=\sum_k\langle a^\dagger_{k,0}a_{k,1}\rangle /L$. Discussing the renormalization of $t_y$ requires a self-consistent solution of the problem, which would, in any case, remain a crude approximation, as it would miss the Luttinger liquid nature of the ground state~\cite{PhysRevB.73.195114}. To provide additional insight, we  evaluate the function $\Omega_{\varphi,U,\rho}$ for the non-interacting problem ($U=0$),  which is equivalent to first order perturbation theory in the interaction $U$: 
\begin{equation}\label{eq:omeganonint}
\begin{split}
\Omega_{\varphi,U,\rho}\simeq \Omega_{\varphi,\rho}=\sum_{s=\pm}\frac\sigma{2\pi} El\left[k^s_F,-\frac{4t_x^2\sin^2(\varphi/2)}{t_y^2}\right]\,,
\end{split}
\end{equation}
where $k_F^{s}$ are the Fermi quasi-momenta of the $s=\pm$ band. $El[k,r]$ is the Elliptic function of the first kind:
\begin{equation}
El\left[k,-\frac1a\right] \equiv \sqrt a \int_0^k \frac{dq}{\sqrt{\sin^2(q)+a}}\,,
\end{equation} 
with the properties $El\left[0,-1/a\right]=0$ and $El\left[k,0\right]=k$. For the experimentally relevant parameters, $\varphi=0.32\pi$, $t_x\approx t_y$ and $\rho<1/2$, $k_F^->k_F^+$ and thus $\Omega_{\varphi,\rho}$ is a positive quantity. Thus, at the mean-field level, repulsive interactions increase the effective transverse coupling between the chains, and stabilize the universal regime for the Hall response corresponding to Eqs.~\eqref{eq:univershall} and~\eqref{eq:rhohexp} in the main text.  

We conclude by stressing that the mean field analysis will in principle change in the situation where the transverse field $E_y\neq0$. However, given the clear limitations of this approximation, we restrain to extend the mean-field analysis to this case and will just consider an effectively increased $t_y$ for the numerical simulations. 

The numerical simulations reproduce the experimental protocol described in the main text. Based on Eq.~\eqref{eq:typ}, we consider a non-interacting theory with an increased transverse hopping. For $t_y^*>5t_x$, the numerical simulations based on the mean-field approximation reproduce with good accuracy the time-evolution of the current $J_x$ and the polarization $P_y$ obtained from more accurate   density
matrix renormalization group (DMRG) \cite{PhysRevLett.69.2863} calculations performed at zero temperature. We invite the reader interested in the details of this comparison to consult the Supplementary Materials of Ref.~\cite{Zhou2023}. We thus consider a value of the renormalized hopping $t_y^*=6t_x$ and extended the simulation starting from a state at finite temperature $T$ in the grand-canonical ensemble. 

\vspace{1em}
\noindent\textbf{\large Supplementary References}

\end{document}